\documentclass[12pt,a4paper]{article}
\usepackage[T1]{fontenc}
\usepackage[utf8]{inputenc}
\usepackage{amsmath,amssymb}
\usepackage[english,russian]{babel}
\usepackage{graphicx}
\usepackage[bottom]{footmisc}

\usepackage[section]{placeins}
\usepackage{float}

\usepackage{graphicx}
\newcommand{\be}{\begin{equation}\label}
\newcommand{\ee}{\end{equation}}
\newcommand{\prt}{\partial}
\newcommand{\p}{\prime}

\usepackage{layout}
\begin{document}
	\selectlanguage{English}
	
	\title{\hrule\vspace{0.1cm} {\small \vspace{0.1cm}}\hrule \vspace{2.0cm}
		{On a Crucial Role of Gravity in the Formation of Elementary Particles }}

\author{ 
	Ahmed Alharthy~\footnote{E-mail:~ahmedal7@protonmail.com}\\
	Kassandrov V.V~\footnote{E-mail:~vkassan@sci.pfu.edu.ru}\\
	{$^{\dagger}$ \small \em Institute of Gravitation and Cosmology}, \\ {\small \em Peoples' Friendship University of Russia}
}
		\maketitle

\begin{abstract}
     We consider the model of minimally interacting electromagnetic, gravitational and massive scalar fields free of any additional nonlinearities. In the dimensionless form the Lagranginan contains only one parameter $\gamma =(m\sqrt{G}/e)^2$ which corresponds to the ratio of gravitational and electromagnetic interactions and, for a typical elementary particle, is about $10^{-40}$ in value. However, regular (soliton-like) solutions can exist only for $\gamma \ne 0$ so that gravity would be necessary to form the structure of an (extended) elementary particle. 
Unfortunately  (in the stationary spherically symmetrical case), the numerical procedure breaks in the range $\gamma \le 0.9$ so that it remains obscure whether the particlelike solutions actually exist in the model. Nonetheless, for $\gamma \sim 1$ we obtain, making use of the minimal energy requirement, a discrete set of (horizon-free) electrically charged regular solutions of the Planck's range mass and dimensions  (``maximons'', ``planckeons'', etc.). In the limit $\gamma \to \infty$ the model reduces to the well known coupled system of the Einstein and Klein-Gordon equations. We obtain -- to our knowledge -- for the first time, the discrete spectrum of neutral soliton-like  solutions (``mini-boson stars'', ``soliton stars'', etc.) \end{abstract}
	
	Keywords: {Soliton-like solutions; gravitational/electromagnetic ratio; boson stars}

	\vspace{0.4cm}
	\noindent

\section{Introduction. Gravity in particle physics}

\hspace{1em} It is generally accepted that gravity's role in the process of formation and structure of elementary particles is negligible, due to the very small intensity of gravitational interaction. Quantitively, the ratio $\gamma$ of Newton's attraction and Coulomb's repulsion forces between two identical charged massive particles (separated by arbitrary distance)  is  
\be{ratio}
\gamma=\frac{e^2_G}{e^2}\equiv(\frac{m\sqrt{G}}{e})^2,
\ee
where $e_G=m\sqrt G$ represents the effective gravitational charge while  $m$ and $e$ are the mass and the (elementary) electric charge of the particle, respectively. For a typical elementary particle $\gamma$ is very small, $\gamma \sim 10^{-40}$ !  Forces become compatible only at the Planck's scale $m\sim 10^{-5} gr$. 

\hspace{1em} However, there are a number of indications on potential - essential- role that gravity can play in the microworld. Some indications come from the structure of the celebrated {\it Kerr-Newman solution} to the Einstein-Maxwell electrovacuum system. Indeed,  for the value of the Kerr parameter $a=\hbar/2mc$ which corresponds to the proper value of the electron's spin, the solution is free of horizon, has a ring-like singularity and can reproduce almost all quantum numbers of the electron~\cite{Burin, Newman}. Moreover, the gyromagntic ratio is equal to that of Dirac's fermion~\cite{Carter, Newman2}, while ``self-quantization'' of the electric charge can be naturally achieved in the framework of {\it algebrodynamic}~\cite{Kassandr}.

\hspace{1em} With that said, we are not aware of any serious attempt to theoretically obtain such an extremely small value of this ``magic'' dimensionless number. In this paper we present a simple field model in which it seems possible to naturally fix a particular value of $\gamma$.

\hspace{1em} In other approaches (see, e.g.,~\cite{Burin2,Dymnik,Arkani}) one also claims to construct a realistic model of an {\it extended} elementary particle in which gravitational self-interaction plays an essential role. Usually, these models are based on everywhere regular solutions ({\it soliton-like, particlelike}, etc.) for a coupled system of the gravitational and other fields' equations. 

\hspace{1em} Even in a model without additionally inserted nonlinearity of the involved fields, corresponding systems of equations are effectively nonlinear. As an example, consider the simple non-relativistic {\it Schr{\"o}dinger-Poisson system of equations} proposed in~\cite{Diosi,Penrose96}~\footnote{ View on the possible fundamental importance and derivation of the non-relativistic Schr{\"o}dinger-Poisson equation ( also known as Schr{\"o}dinger-Newton equation) from the coupled Einstein-Dirac (or Klein-Gordon) system of equations may be found in a large number of works (see, e.g.,~\cite{Bahrami,Hu,Guilini})},  as a model for describing the process of the wavefunction reduction induced by self-gravitation of distributed parts of a massive quantum-like particle, 
\be{Poisson}
-\frac{\hbar^2}{2m}\Delta \psi +U\psi = E\psi, ~~\Delta U = 4\pi Gm^2 \vert \psi\vert^2,  
 \ee
in which the gravitational self-energy $U$ which defines the particle's $\psi$-function is itself determined by the probability distribution $\vert \psi \vert^2$. 
Despite the fact that both equations are linear by themselves, after resolving the second equation w.r.t $U$ and substituting $U$ into the first equation, we obtain a highly nonlinear equation of the form
\be{Schrod}
\frac{\hbar^2}{2m}\Delta (\frac{\Delta\psi}{\psi})  =  4\pi Gm^2 \vert \psi\vert^2
\ee 
which is in the fourth order in derivative. \\

\hspace{1em} In~\cite{Tod} a discrete energy spectrum of regular solutions to (\ref{Poisson}) has been obtained in the stationary spherical symmetric case. Note that relativistic generalization of this model~\cite{IJMPA} is based on a coupled system of Dirac and Maxwell-like equations in which the latter effectively substitute the Einstein gravity.   

\hspace{1em} As a rule, corresponding objects must be treated as {\it macroscopic} since their mass are in the Planck's range or even greater~\cite{Torres}.  On the other hand, they differ from the so-called {\it regular black holes} (see, e.g.,~\cite{Bronnik}) being not only free of the central singularity but free of any horizon as well. In~\cite{Dymnik2} these have been called {\it gravitational lumps} (G-lumps), in~\cite{Friedberg} -- {\it (mini-) soliton stars}. In the case when the scalar field is an essential constituent of the model, such field distributions are called {\it (mini-) boson stars}~\cite{Mielke, Jetzer}. 

\hspace{1em} According to one of the classifications presented - in~\cite{Torres}- in the case of scalar field coupled to gravity,  one distinguishes between mini-boson stars, boson stars and (mini)-soliton stars in the following way: 

  \begin{itemize}
  \item
 It is a mini-boson star if the only non-kinetic term present in the scalar field Lagrangian is the mass term.
   
 \item
 It is a boson star if additional nonlinearity is inserted into the Klein-Gordon equation.

\item 
 (Mini)-soliton stars correspond to solutions which have a flat limit, i.e. to the soliton-like solutions that exist in the Minkowski space-time.                       

 \end{itemize}

\hspace{1em} As to the range of dimensions and mass corresponding to these objects, they depend not only on the type of solution considered but on the range of parameters selected, namely, on the value of the bare mass $m$ and on the scalar self-interaction constant. That's why various solutions can be identified with quite different (astro-)physical objects.

\hspace{1em} Such models are semi-classical in origin, since what we seek for are regular (soliton-like) solutions to the field equations which possess finite and, sometimes, {\it discrete} spectrum of Noether's integrals defining the values of corresponding quantum numbers (mass, spin, charge, etc.). Thereby, 
one usually disregards gravity and deals with nonlinear fundamental fields in Minkowski space-time. Gravity can be then taken into account as an extremely small correction. 

\hspace{1em} A major difficulty in such models is the form of the chosen nonlinearity, which remains almost arbitrary.  
A proposed alternative to eliminate the "arbitrariness" in the form of the nonlinearity, is to compose a system where the well-known set of {\it linear fundamental fields} (corresponding to Maxwell, Dirac or Klein-Gordon equations) are coupled, and thus form effectively nonlinear system of equations. By that, the form of effective nonlinearity becomes rigidly fixed by the requirement of {\it gauge invariance}.

      
\section{Rosen's model and its counterpart on a curved manifold}

\hspace{1em} From various models for extended particles, the Dirac-Maxwell system of equations for the spinor and electromagnetic fields coupled via the minimal electromagnetic interaction is, certainly, one of the most attractive. Remarkably, it is a direct {\it semi-classical analogue of the operator equations of quantum electrodynamics}. As it was discovered in~\cite{Wakano, TerlKass, Cooperstock},  the model possesses a class of regular stationary axisymmetric solutions which correspond to {\it fermions of any half-integer spin}~\cite{TerlKass} with an elementary electric charge and proper value of the magnetic moment. Unfortunately, for all the solutions, {\it the masses turn out to be negative}. 

\hspace{1em} A much simpler model of minimally coupled complex Klein-Gordon ($\varphi$) and Maxwell ($A_\mu$) fields was first proposed by N. Rosen as early as in 1939~\cite{Rosen}. \\The Lagrangian reads ($F_{\mu\nu}:=\prt_\mu A_\nu - \prt_\nu A_\mu,~~ D_\mu:=\prt_\mu - ieA_\mu$): 
\be{RosenLagr}
L=-\frac{1}{2} F^{\mu\nu} F_{\mu\nu}- D^*_\mu \phi^* D^\mu \phi + m^2 \phi^* \phi,      
\ee
 For the field equations corresponding to (\ref{RosenLagr}),
 there exists a class of regular stationary spherically symmetrical solutions which can possess  {\it any integer spin}~\cite{EdjoKassTerl}. Rosen had even made an attempt~\cite{Rosen2} to extract the value of the {\it fine structure constant} $\alpha$ making use of the solution corresponding to the minimum of the self-energy. However, the attempt had failed and the masses of all the solutions, again, turned out negative.      

\hspace{1em} As an attempt to solve the problem of negative mass, one can try to change the signs of the kinetic and mass terms in the Lagrangian. In the flat case, this leads to the canonical Lagrangian of the Klein-Gordon field and the solutions with a  positive definite energy density.  However, it is easy to demonstrate that {\it regular} solutions do not exist in this case, even if minimal interaction with Maxwell's field was taken into account, as in (\ref{RosenLagr}). 

\hspace{1em} At this point it seems quite natural to include gravity, in order to maintain the {\it regular} solutions. Indeed, even with such a small interaction constant, the behavior of field functions  can change drastically and allow for the existence of regular solutions with positive proper energy (mass). This mean that, according to the above described classification~\cite{Torres, Friedberg}, the solutions will not belong  to the so-called soliton-like class but will describe an (electrically charged) ``mini-boson'' distributions. However, as we have already mentioned, this and other classifications are rather "voluntaristical".        

\hspace{1em} Thus, we are led to consider the following {\it 3-field Lagrangian}:
\be{OurLagr}
L = -\frac{1}{16\pi}F_{\alpha\mu}F_{\beta\nu} g^{\alpha\beta}g^{\mu\nu} +\frac{1}{8\pi}D^*_\mu \phi^* g^{\mu\nu}D_\nu \phi -\frac{k^2}{8\pi} \phi^*\phi + \frac{c^4}{16\pi G} R,
\ee
where $D_\mu:=\prt_\mu-i\epsilon A_\mu$, $F_{\mu\nu}:=\prt_\mu A_\nu - \prt_\nu A_\mu$ and $R$ being the Riemannian curvature scalar. The dimensional constants $\epsilon$ and $k$ (or, equivalently, $m=\hbar k/c$) should be fixed {\it a posteriori}. However, in order to establish the right ``charge-spin'' ratio~\cite{EdjoKassTerl}  and in correspondence with quantum theory, we set below  $\epsilon:=e/\hbar c$, and it would be also natural to consider the priming mass $m$ in the range typical for elementary particles.


\section{General characteristics of the 3-field model}

\hspace{1em} After scale transformations of coordinates (reduction to the Compton scale), and field functions of the form: 
$$
x_\mu \mapsto \frac{\hbar}{mc} x_\mu,~~ A_\mu \mapsto \frac{mc^2}{e} A_\mu, ~~\phi \mapsto \frac{mc^2}{e} \phi,  
$$
the Lagrangian (\ref{OurLagr}) transforms in the following way: 
$$
L \mapsto \frac{mc^2}{\alpha} k^3 L,   
$$
where $\alpha:=e^2/\hbar c$ is the fine structure constant, and the dimensionless Lagrangian $L$ has the following form ($D_\mu:=\prt_\mu-i A_\mu$):
\be{LagrDL}
L = -\frac{1}{16\pi}F_{\mu\nu}F^{\mu\nu} +\frac{1}{8\pi}D^*_\mu \phi^* D^\mu \phi -\frac{1}{8\pi} \phi^*\phi + \frac{1}{16\pi \gamma} R. 
\ee  
$L$ contains only a sole {\it dimensionless} parameter, $\gamma:=(m\sqrt G / e)^2$ -- as discussed earlier --  being 
of the order $10^{-40}$ for a typical elementary particle. However, {\bf we cannot set it zero since regular solutions disappear in this case} !


\hspace{1em} Exploiting regular solutions of the 3-field model for describing the structure of (charged massive) elementary particles seems quite natural. Indeed, any of these particles produce an electromagnetic, gravitational and a wave-like fields. On the other hand, the choice of the Klein-Gordon field can be considered only as a first step to the study of the more attractive but complicated {\it Dirac-Maxwell-Einstein model} .

\hspace{1em} Even in the case of scalar field, when the field equations allow for spherically symmetric ansatz, the search for regular solutions is an extremely complicated problem. We are aware of only one work~\cite{EdjoTerl} where this problem had been posed and preliminary results obtained. However, the paper is not complete and contain some imprecise results in numerical calculations. Therefore, we have made an attempt here to further elaborate the above 3-field model.   

\hspace{1em} Varying the action functional $S=\int L \sqrt{-g} d^4 x$ corresponding to  Lagrangian (\ref{LagrDL}) w.r.t $\phi^*, A_\mu$ and the metric $g^{\mu\nu}$, we obtain the field equations in the form
\be{eqs}
\begin{array}{ccc}
\frac{1}{\sqrt{-g}} D_\mu (\sqrt{-g} g^{\mu\nu} D_\nu \phi) = \phi,   \\
\frac{1}{\sqrt{-g}} \prt_\nu (\sqrt{-g} F^{\mu\nu}) = \frac{i}{2}(\phi^* D^\mu \phi - D^{*\mu} \phi^* \phi), \\
R_{\mu\nu} - \frac{1}{2} g_{\mu\nu} R =\gamma {\tilde T}_{\mu\nu}, 
\end{array}
\ee    
where $g=\det\vert g_{\mu\nu}\vert$ and  $\tilde T_{\mu\nu}$ is the (renormed) energy-momentum tensor, 
\be{emt}
\frac{1}{2}\tilde T_{\mu\nu}=-F_{\mu\alpha}F_{\nu\beta} g^{\alpha\beta} +\frac{1}{4} g_{\mu\nu} F^2+D^*_\mu \phi^* D_\nu \phi +D^*_\nu \phi^* D_\mu \phi - g_{\mu\nu}(D^2  - \phi^*\phi), 
\ee
$D^2$ being the invariant $D^2:=D^*_\alpha \phi^* g^{\alpha \beta} D_\beta \phi$.
 
\hspace{1em} Finally, from the corresponding equations (\ref{eqs}), one can obtain the dimensionless (electric charge $q$, and energy $w$) of the regular solution which are related to corresponding dimensional (electric charge $Q$, energy $W$ ) by :
\be{chargemassdim}
Q=(e/\alpha) q, ~W=(mc^2/\alpha) w,
\ee
so that {\bf one can ensure their values be typical for elementary particles if for $\gamma \approx 10^{-40}$ one would have for dimensionless charge and energy (mass) $q = \alpha \approx w$}.


\section{Stationary spherically symmetric ansatz}

\hspace{1em} Equations (\ref{eqs}) allow for the following canonical ansatz:
\be{anzatz}
\begin{array}{cc}
ds^2 = f^2 dt^2 - h^{-2} dr^2 -r^2 (\sin^2 \theta d\theta^2 +d\varphi^2), \\
\phi = \varphi(r) e^{-i\omega t}, ~~A_\mu = \{\Phi(r), \bf 0\},
\end{array}
\ee
and the two metric functions~\footnote{To avoid {\it radicals} in the equations we define metric functions via {\it squares}. Below we shall see that for any $r$ and any solution both of these are positive} $f^2(r), h^2(r)$ depend on the radial coordinate $r$.

\hspace{1em} Corresponding  {\it action potential} takes the form:
\be{action}
S =\frac{1}{2\gamma} \int dr f (\frac{1}{h} - h - 2 h^\p r)+\frac{1}{2}\int r^2 dr  (\chi^{\p2} \frac{h}{f} - \varphi^{\p2} f h +\chi^2 \varphi^2 \frac{1}{fh} - \varphi^2 \frac{f}{h}), 
\ee
where $(\p)$ denotes derivation w.r.t the coordinate $r$.  Varying (\ref{action}) w.r.t $\varphi(r), \Phi(r), f(r), h(r)$ one obtains the field equations in the form
 \be{spheqs}
 \begin{array}{cccc}
 \varphi^{\p\p} +\frac{2}{r} \varphi^\p + \varphi^\p (\frac{h^\p}{h} + \frac{f^\p}{f})  = \frac{1}{h^2}\varphi(1-\frac{\chi^2}{f^2}), \\
 \chi^{\p\p} +\frac{2}{r} \chi^\p + \chi^\p (\frac{h^\p}{h} - \frac{f^\p}{f})  = \frac{1}{h^2}\chi \varphi^2, \\
 1-h^2 - (h^2)^\p r = \gamma r^2 (\chi^{\p2} \frac{h^2}{f^2} + \varphi^{\p2} h^2 +\chi^2 \varphi^2 \frac{1}{f^2} + \varphi^2), \\
 1-h^2 - 2(\frac{f^\p}{f}) h^2 r = \gamma r^2 (\chi^{\p2} \frac{h^2}{f^2} - \varphi^{\p2} h^2 -\chi^2 \varphi^2 \frac{1}{f^2} + \varphi^2),
  \end{array}
 \ee 
where the shifted electric potential is $\chi:=\Phi + \omega$. The last two equations in (\ref{spheqs}) correspond to two independent Einstein equations corresponding to the ($tt$) and ($rr$) components, respectively. From the third equation of (\ref{spheqs}), the energy density is seen to be positive definite.

\hspace{1em} As for the other two Einstein equations which are not contained in the system (\ref{eqs})  (correspond to ($\theta \theta$) and ($\phi \phi$) components), they both have the same form: 
\be{EinstAdd}
- (r h)^2 (\frac{f^{\p \p}}{f}
+\frac{f^\p}{f}\frac{h^\p}{h}+\frac{1}{r}(\frac{f^\p}{f}+ \frac{h^\p}{h}))=
\gamma r^2 (-\chi^{\p2} \frac{h^2}{f^2} +\varphi^{\p2} h^2 -\chi^2 \varphi^2 \frac{1}{f^2} + \varphi^2),
\ee
and are known to identically hold on the solutions to (\ref{spheqs}). It is worthy to note that according to the stress-energy tensor (SET) components in the r.h.s of (\ref{spheqs}) and (\ref{EinstAdd}), corresponding radial and tangential pressure are generally different. However, as we shall see below, for regular solutions near the center $r\to 0$ the derivative terms therein turn to zero, so that asymptotically, one has a fully isotropic SET and corresponding {\it de Sitter core}.  

\hspace{1em} We now start to seek for regular asymptotically flat solutions to the system (\ref{spheqs}). 
Specifically, at great distances we should achieve (by proper fitting of the initial values) 
\be{asymp}
\varphi \sim \exp({- \sqrt{1-\omega^2}~r}), ~ \chi \sim \omega + q/r, ~q<0,
\ee
and for metric functions -- the Reissner - Nordstr{\''o}m asymptotics
\be{asymmetr}
h^2 \sim 1-\frac{M}{r}+\gamma \frac{q^2}{r^2}, ~~f^2 \sim  1-\frac{M}{r}+\gamma \frac{q^2}{r^2},
\ee 
$M$ being  (twice the) {\it gravitational charge} of the distribution.
 
\hspace{1em} Near the center $r\to 0$ regular behavior of field functions has the form:
\be{asymporg}
\begin{array}{cccc}
\varphi\sim a(1-(B^2 - 1)r^2+...), \\
\chi \sim \delta (1+\frac{a^2}{6} r^2 + ...) \\
h \sim 1 -\gamma \frac{a^2}{6} (B^2+1) r^2 + ...\\
f \sim A (1+\gamma \frac{a^2}{6}(2B^2 - 1)r^2 + ... )
\end{array}
\ee
in which a very important parameter $B:=\delta/A$  does enter. $B$ is the ratio of  initial values $\delta=\chi(0)$ and  $A=f(0)$.   
    
\hspace{1em} Another thing to notice is that the system (\ref{spheqs}) is covariant under the transformation:
\be{symm}
\chi \mapsto K \chi, ~~f \mapsto K f, ~~K=const
 \ee           
which shows that the parameter $B$ preserves its value and thus, can be accepted as the principal parameter defining the form and characteristics of regular solutions to (\ref{spheqs}). Note that the symmetry (\ref{symm}) is related to the possibility of redefining the time coordinate and thus, of the frequency parameter $\omega$ , which enters the expression for the shifted electric potential  $\chi= \omega +\Phi(r)$.  

   
\section{ Regular solutions: procedure of numerical integration}
\hspace{1em} Let us now consider the procedure of integrating the system (\ref{spheqs}),  seeking the solutions with the above described regular behavior in the center and at infinity. For a given $\gamma$, and for any selected value of the sole free parameter $B$, one takes arbitrary $\delta=\chi(0)$ (and thus corresponding $A=f(0) = \delta/B$) and starts to fit $a=\varphi(0)$ to ensure~\footnote{For this, we use the well-known ``bracketing'' method, see, e.g.,~\cite{Tod,Hydrogen}}  exponential decrease of $\varphi$ at sufficiently large $r$, up to the value $r=r_0$ at which the solution either blows up, $\varphi^\p (r_0)>0$, or change its sign, $\varphi(r_0) <0$. 

\hspace{1em} From the last equation (\ref{spheqs}) it follows  that in the range $r>r_0$ 
the asymptotic behavior of metric function $f(r)$ will be actually more general than given in (\ref{asymmetr}), and it will be non-Halileian, namely, 
\be{asymgen}
f^2 \sim C (1-\frac{M}{r}+\gamma \frac{q^2}{r^2}),
\ee
with $C$ being an arbitrary constant. However, using the symmetry (\ref{symm}) we can {\it renorm} the obtained solution by shifting its asymptotic value to the Halileian one represented in (\ref{asymp}). Typical forms of the scalar and shifted electric potential fields with the metric functions  (for the values of parameters $\gamma=1, B=3.5$ which correspond to the local minimum of energy, see below) are depicted in Figs. 1-2 and Figs. 3-4, respectively. We see that the metrics does not change its signature so that  {\bf all the obtained solutions are free of horizon}. We note that in the neutral case, when the  electromagnetic field is absent, such property (freedom of horizon) had been first discovered in ~\cite{Kaup} and analytically proved  later in~\cite{Bronnik2} (called  the ``no-go theorem'' therein).

\hspace{1em} Now, using values of the functions at the breaking point $r=r_0$ and asymptotics (\ref{asymp}), (\ref{asymmetr}),  we can compute the characteristics of the obtained regular solution, namely the electric charge $q=-r_0^2 \chi^\p(r_0)$,  frequency $\omega=\chi(r_0)+r_0 \chi^\p(r_0)$ and gravitational charge $M = 2r_0 (1 - h^2(r_0) -\frac{1}{2} r_0 h^{2 \p}(r_0))$. 

\hspace{1em} Finally, using the first and third equation in (\ref{spheqs}), we integrate (in the range $r=0 ... r_0$) the charge and energy density with the already found initial values of field functions and obtain the total electric charge $\tilde q$ 
$$
\tilde q = - \int r^2 dr  \frac{\chi \varphi^2}{hf}
 $$    
 and the inertial mass $\tilde M$, 
 $$
 \tilde M =\int r^2 dr (\chi^{\p2} \frac{h^2}{f^2} + \varphi^{\p2} h^2 +\chi^2 \varphi^2 \frac{1}{f^2} + \varphi^2).
 $$
 
\begin{figure}[!htb]
	\begin{center}    
	
    \includegraphics[width=0.7\textwidth]{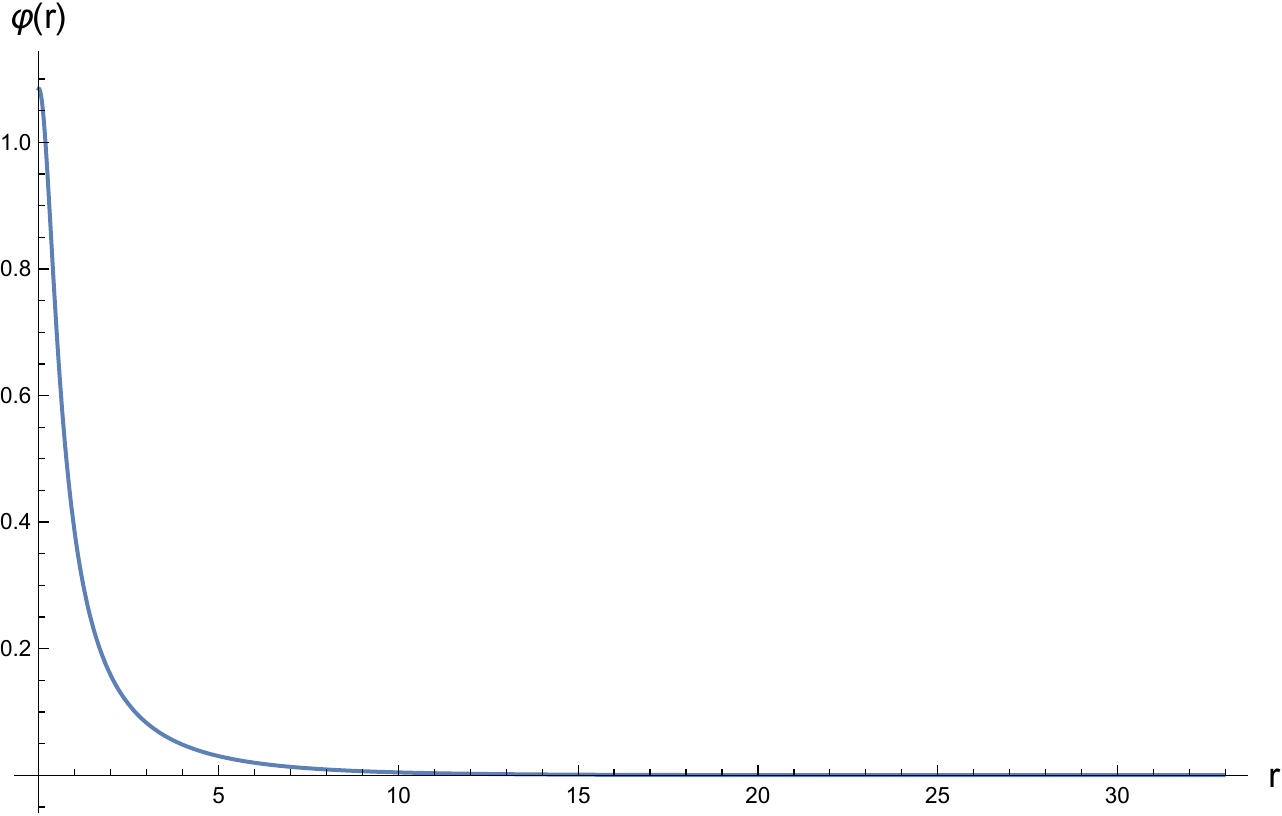}
    \caption{Typical form of the $\varphi(r)$ distribution exponentially decreasing at large $r$}
   	\label{fig:image3}
	\end{center}  
\end{figure}
	
	\begin{figure}[!htb]
	\begin{center}    
    \includegraphics[width=0.7\textwidth]{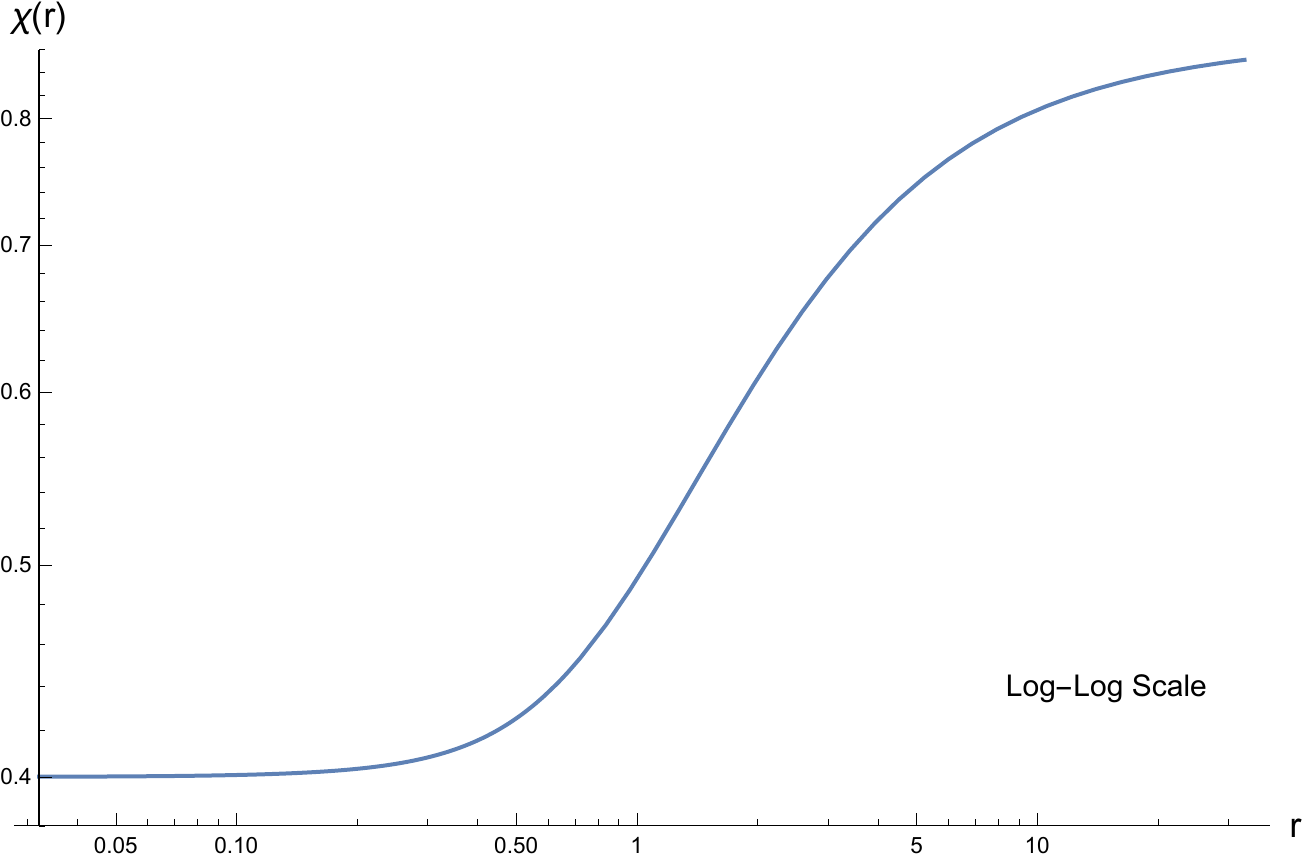}
    \caption{Typical form of the (shifted) electrostatic potential $\chi(r)$, $\chi(\infty)=\omega$}
   	\label{fig:image3}
	\end{center}  
	\end{figure}
	
	\begin{figure}[!htb]
	\begin{center}    
    \includegraphics[width=0.7\textwidth]{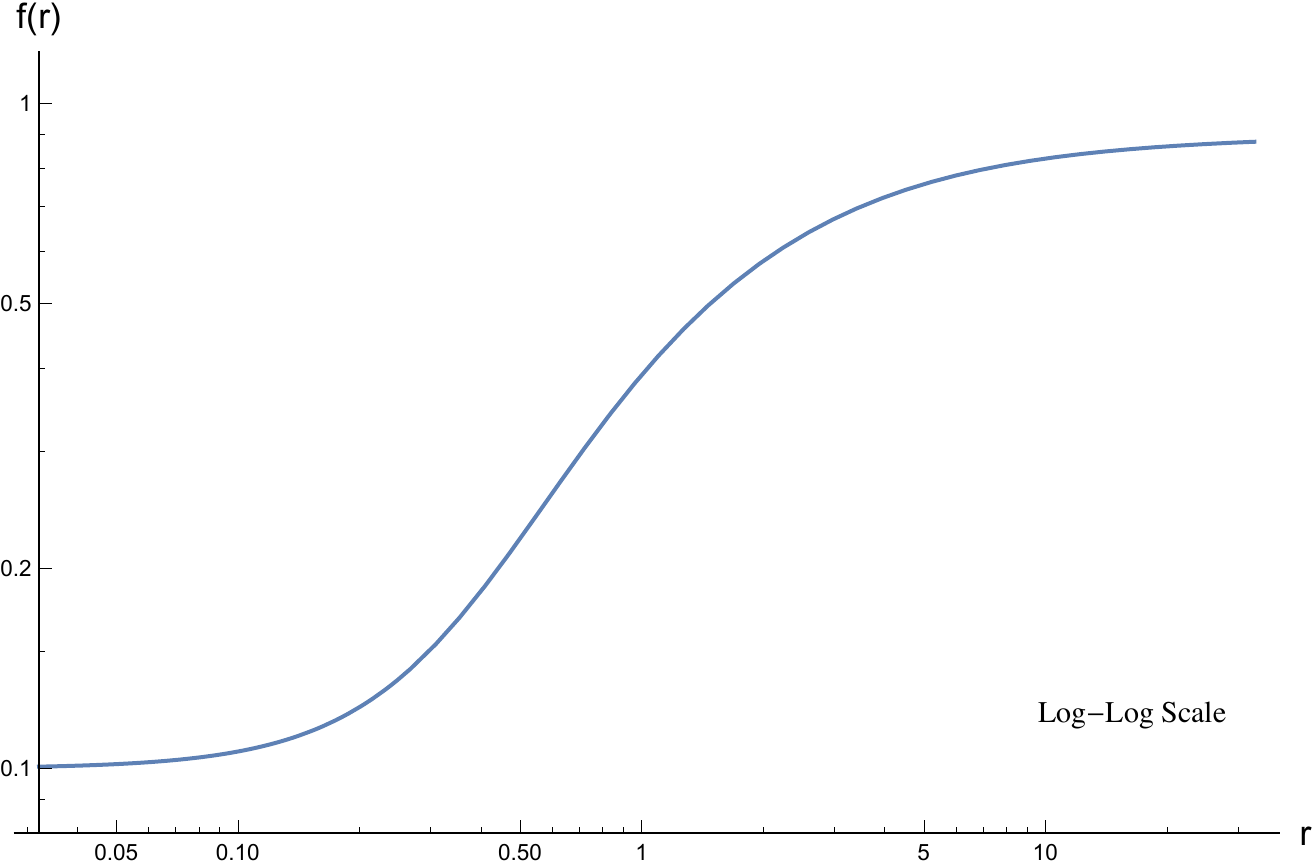}
    \caption{Typical form of a (horizon-free) metric function $\sqrt{g_{tt}} =f(r)$; note that $f(0) \ne 0,\\ f(\infty) = 1$}
   	\label{fig:image3}
	\end{center}  
	\end{figure}

\begin{figure}[!htb]
	\begin{center}    
    \includegraphics[width=0.7\textwidth]{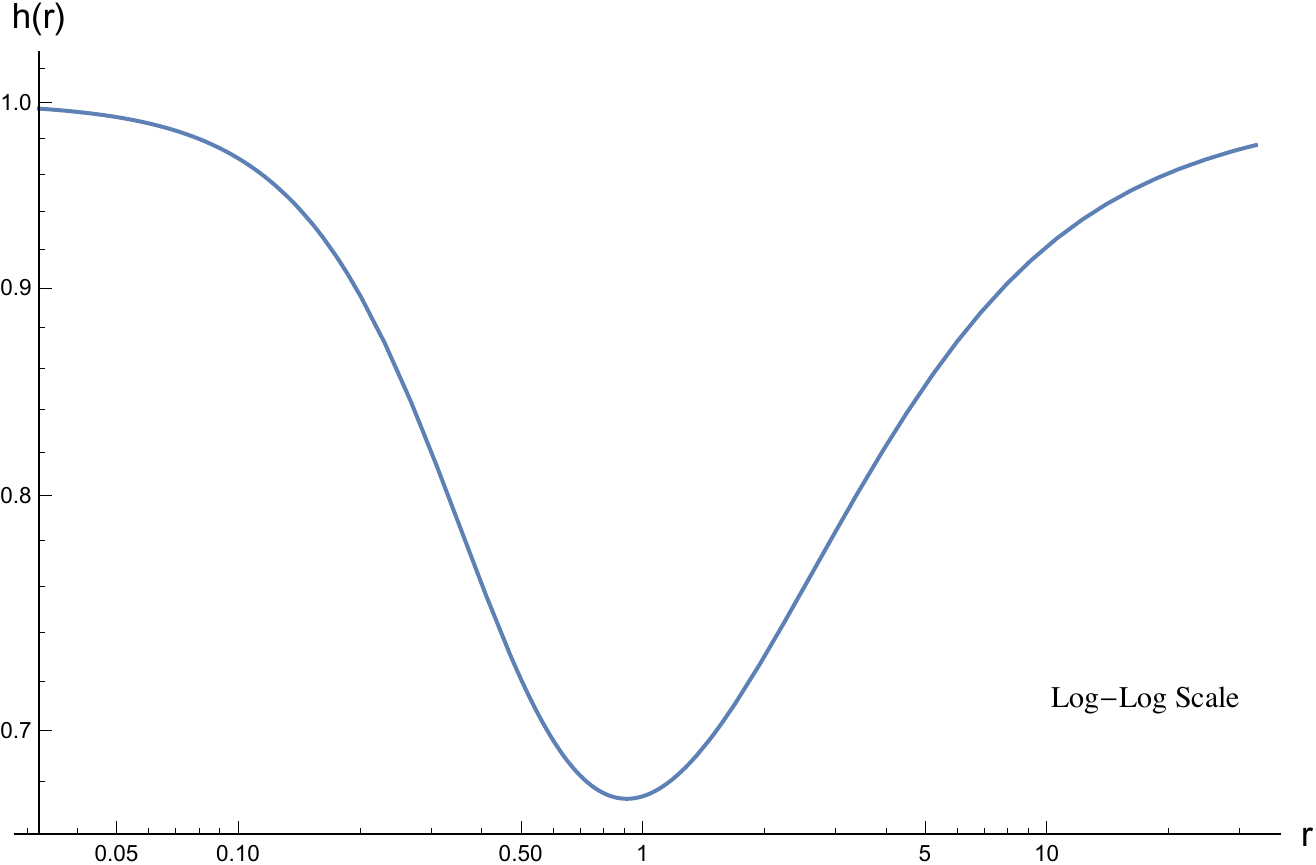}
    \caption{Typical form of a (horizon-free) metric function $\frac{-1}{\sqrt{g_{rr}}} =h(r)$ ; note that \\ $h(0)=h(\infty)=1$}
   	\label{fig:image3}
	\end{center}  
\end{figure}

\hspace{1em} If the precision of integration is sufficient one should satisfy the integral identities 
$$
\tilde q = q, ~~~\gamma \tilde M =M,
$$
where the mass equation has the meaning of the {\it equivalence principle}. 
  
 
  \clearpage
\section{Regular solution: characteristics and dependence on parameters}

\hspace{1em} We present now the results of numerical study of the regular solutions to (\ref{spheqs}). For the value of parameter\footnote{The solution with $\gamma =1$  for a particular value of $\omega$ had been previously obtained in~\cite{EdjoTerl}} $\gamma =1$  the plots of electric charge $q$, mass $M$ and frequency $\omega$ associated with the solutions are plotted in Figs. 5-7 as functions of the principal free parameter $B$. 
\begin{figure}[h!]
	\begin{center}    
    \includegraphics[width=0.7\textwidth]{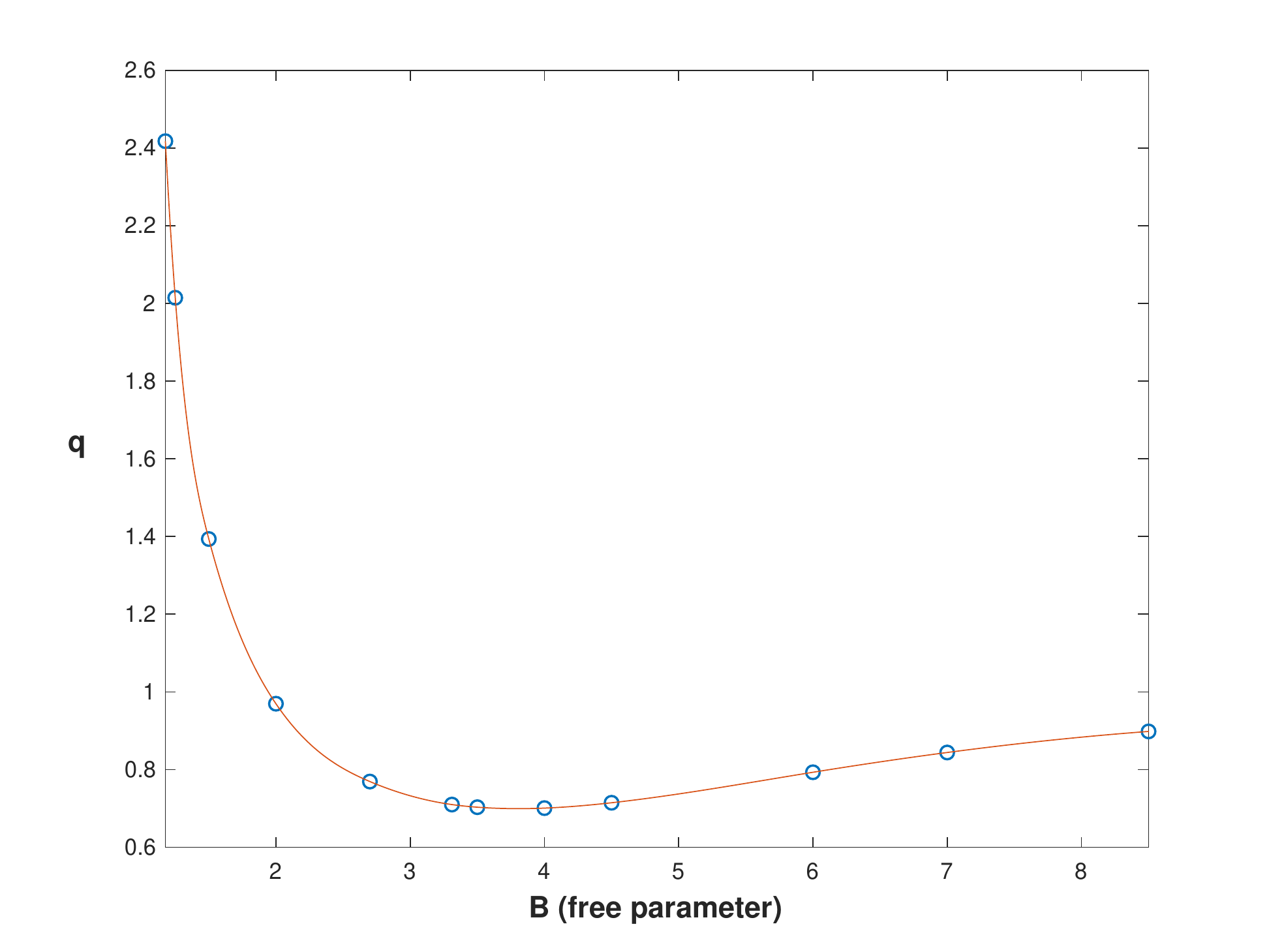}
    \caption{Electric charge $q$ against the free parameter $B$}
   	\label{fig:image3}
	\end{center} 
\end{figure}
	
\begin{figure}[H]
	\begin{center}    
    \includegraphics[width=0.7\textwidth]{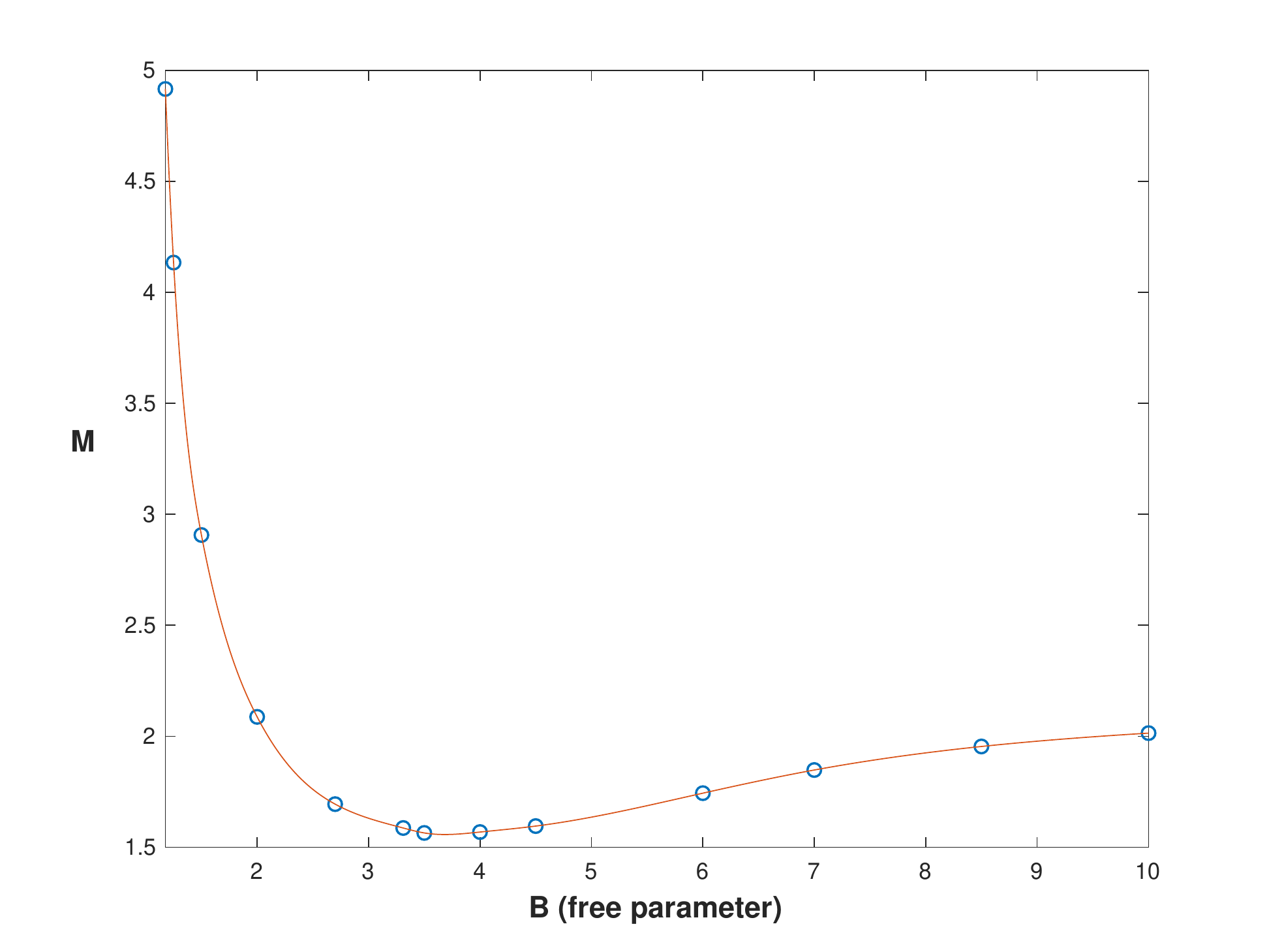}
    \caption{Proper energy (mass) $M$ against the free parameter $B$} 
   	\label{fig:image3}
	\end{center} 
\end{figure}

\begin{figure}[H]
	\begin{center}    
    \includegraphics[width=0.7\textwidth]{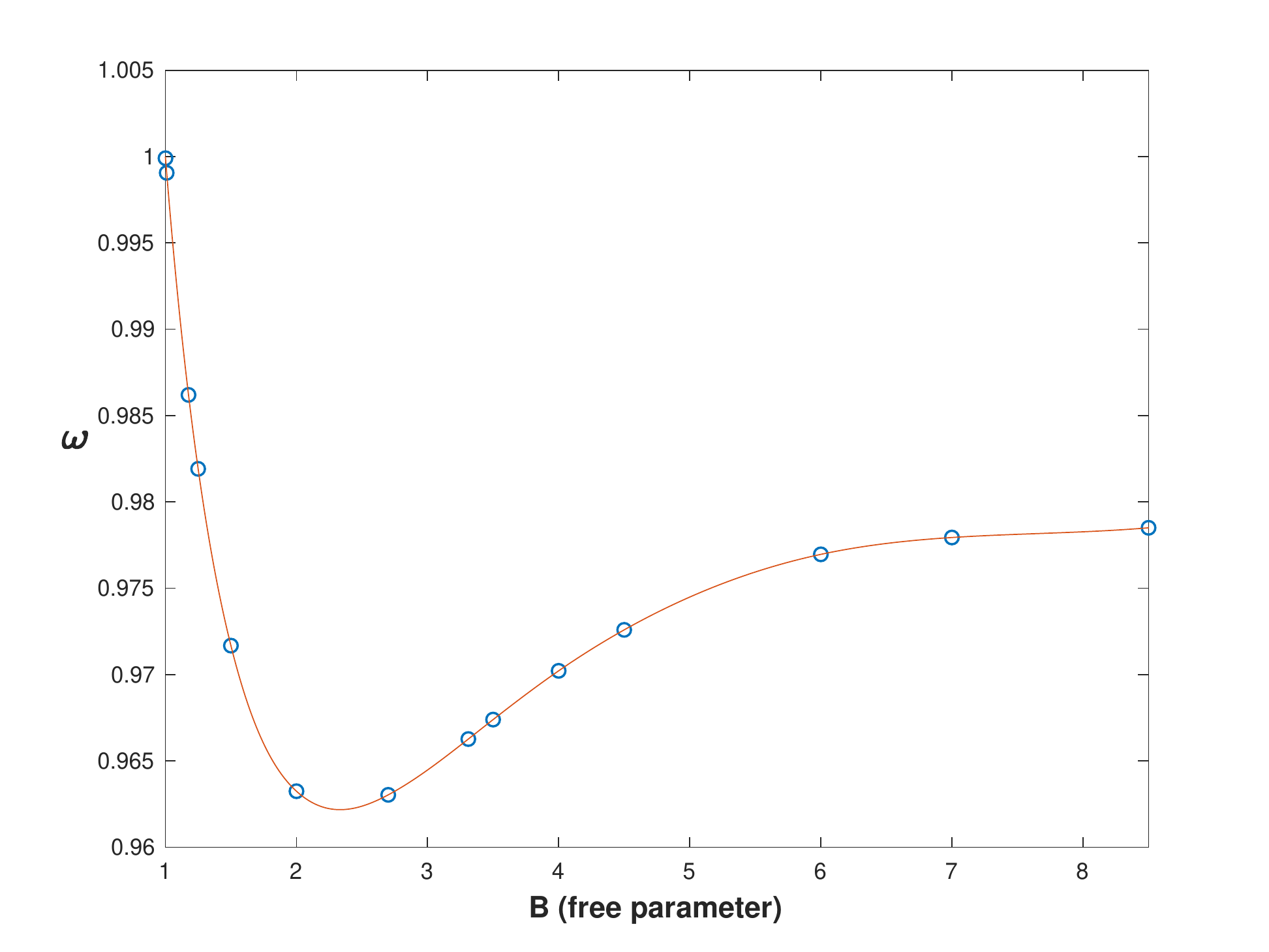}
    \caption{Frequency $\omega$ against the free parameter $B$}
   	\label{fig:image3}
	\end{center}  
\end{figure}

\newpage

\hspace{1em} From these data we conclude that: 
  \begin{itemize}
  \item
  The solutions exist only in a narrow range of the frequency parameter $\omega\sim 1$ near the boundary value $\omega =1$.  For $\omega =1$, the asymptotic of the scalar field differs from the Yukawa's form and corresponds \footnote{In the flat case such ``extreme'' solutions have been discovered in~\cite{TerlKass}} to $\varphi \sim \exp(-\sqrt{8qr~} )$.     
  
 \item
All three characteristics $q,M,\omega$ have a minimum at almost equal values of the free parameter $B$. 

\item 
There exist some additional local minima for the values of electric charge and energy  
(in other range of the parameter $B$, not represented in the figures).  

 \end{itemize}

\hspace{1em} Let us now take into account that a fixed value of $\gamma$ corresponds, in view of (\ref{ratio}), to the bare mass $m=\sqrt{\alpha} M_{Pl}$, where $M_{Pl}$ is the Planck's mass ($M_{Pl}\approx 10^{-5} gr$) , and  $\alpha$ is the fine structure constant. 
For the dimensional mass $M_{dim}$, according to  (\ref{chargemassdim}) , one has ~\cite{EdjoTerl}
\be{Mdim} 
 M_{dim}=\frac{\sqrt{\gamma} M}{\sqrt{\alpha}} M_{Pl}, 
  \ee
  while for the electric charge, 
  \be{Qdim}
  Q_{dim} = \frac{q}{\alpha} e, 
  \ee

\hspace{1em} We assume now that the {\it stable} solution corresponds to the local minimum of the proper energy (mass) which occurs at $B\approx 3.5$ and corresponds to the following characteristics: $q\approx 0.7033,~ M\approx 1.565,~ \omega \approx 0.9674$ and the effective radius of the scalar field distribution is about $R\sim 100$.  Dimensional physical characteristics (\ref{Mdim}), (\ref{Qdim}) then equal to 
 \be{MQdim}
 M_{dim}\approx 18.3 M_{Pl}, ~~Q_{dim}\approx 100 e, ~~R_{dim}\approx \approx 1200 L_{Pl},
 \ee
$L_{Pl} \approx 10^{-33} cm$ being the Planck's length. 

\hspace{1em} Thus, we deal here with an extremely  compact  electrically charged object of a Planck's range mass (hypothetical ``maximon''). However, this object possesses a discrete energy spectrum and is, therefore, ``semi-quantum'' in origin proving the old idea of M.A. Markov on its nature as an elementary particle of a maximal possible mass~\cite{Markov}.     

\hspace{1em} We can further seek for regular solutions which correspond to a number of {\it nodes} for the scalar field function $\varphi(r)$.  For such solutions, there exists a range of the parameter $B$ corresponding to the local minima of the energy. We evaluate the values of minimal energy (mass) of the 1-node and 2-node ``excited''  states as $M^{(1)} \approx 11.19,~~M^{(2)} \approx 28.01$ (compare with $M^{(0)} \approx 1.565$ for the ``ground state'', see above).

\hspace{1em} Let us follow now the dependence of solutions on the parameter $\gamma$. For $\gamma=0.945$, say, the characteristics $q,M,\omega$ on $B$ preserves all the properties represented above for $\gamma=1$. In particular, the minimal energy is now equal to $M\approx 1.897$ and corresponds to  $B\approx 3.4$.     

\hspace{1em} Unfortunately, {\bf the procedure of numerical integration becomes unstable for $\gamma\le 0.9$ so that the most interesting range of parameters corresponding to the characteristics of particles remain unattainable}.  Our attempt to evaluate probable characteristics at such values of $\gamma$ was unsuccessful, even after applying the variational methods. 

\hspace{1em} On the other hand, this means that there still remains a possibility for such particlelike solutions to exist in the model.  It can be conjectured that in this case {\bf one could fix the values of the two free parameters $\gamma$ and $B$ through the requirement of minimum for the values of both the electric charge and energy (mass)}. At the moment, realization of this program (which claims, in particular, to obtain the magic number at the range of $10^{-40}$) looks hardly probable.


\section{Electrically neutral solutions}

\hspace{1em} Consider now the ``macroscopic'' sector of regular solutions which corresponds to the range $\gamma >1$. We obtain from solutions in this range dimensionless electric charge $q$ and mass $M$, which depend weakly on $\gamma$. According to (\ref{Mdim}), that mean physical mass grow as $\sqrt{\gamma}$. For $\gamma >>1$ corresponding solutions can thus be identified as  
(weakly charged, see (\ref{Qdim}))  {\it star-like} or even {\it supermassive} objects.    

\hspace{1em} In the limit $\gamma \to \infty$ which corresponds to $e \to 0$ the model reduces to the system of the minimally coupled Einstein and Klein-Gordon equations derived from the Lagrangian (compare with (\ref{OurLagr}))
\be{ScLagr}
L = \frac{1}{8\pi}\prt_\mu \phi^* g^{\mu\nu}\prt_\nu \phi -\frac{k^2}{8\pi} \phi^*\phi + \frac{c^4}{16\pi G} R,
 \ee          
which in the dimensionless form {\it contains no free parameters at all}. Physical mass can be then evaluated as follows~\cite{Kaup} :
$$
M_{dim} = \frac{M_{Pl}^2}{m} M, 
$$
 with $M$ being the dimensionless mass of the solutions. Thus, for the quantum mechanical range of the bare mass $m$ one deals here with electrically neutral objects with mass $M_{dim}\sim 10^{21} gr \approx 10^{-9} M_{\odot}$, that is, with the hypothetical {\it mini-boson stars} (see, e.g., ~\cite{Friedberg, Jetzer}).  
 
\hspace{1em} In fact, regular solutions to the equations resulting from (\ref{ScLagr}) were considered in several works. Particularly, in~\cite{Kaup} the absence of a horizon for these solutions was discovered. However, the dependence of the characteristics on the sole free parameter $A=f(0)$ which can fix the discrete energy spectrum, to our knowledge, has not yet been retraced.

\hspace{1em} In our model, the dependence of the mass $M$ and the frequency $\omega$ on the parameter $A$ are represented in Figs. 8, 9 respectively. One of the two local minima of energy corresponds to $A\approx 0.16$ and in the minimum $M$ equal to $0.6811$; the other, a deep minimum is achieved when $A$, together with $\omega$, approach the extreme allowed value equal to unity; minimal value of energy is then about $M\sim 0.161$. Note also that the frequency values are located again in a narrow interval $\omega=0.77 ... 0.999$ near the cut-off value $\omega = 1$.

\begin{figure}[h!]
	\begin{center}    
    \includegraphics[width=0.7\textwidth]{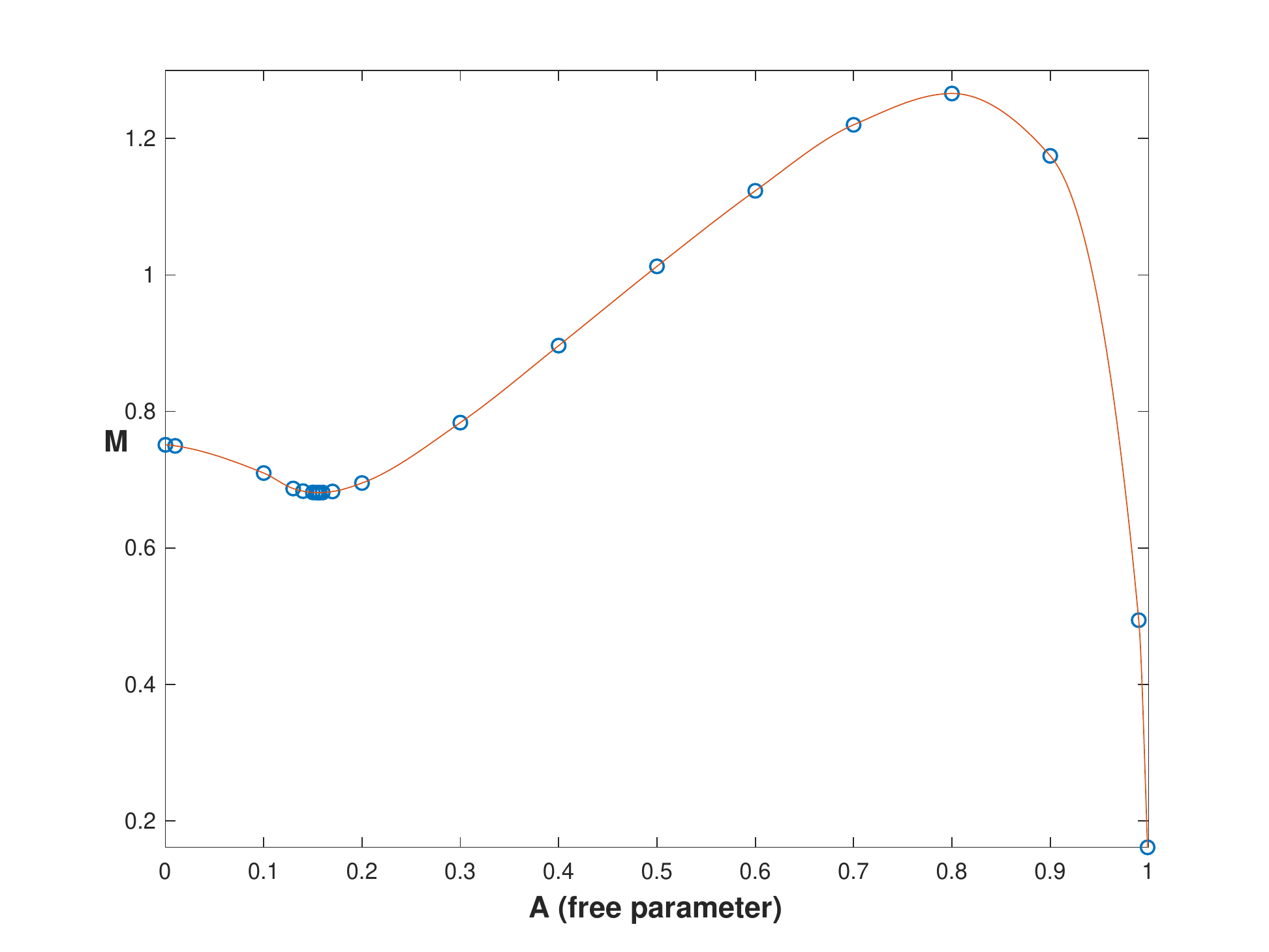}
    \caption{Proper energy (mass) $M$ against the free parameter $A=f(0)$ for the neutral solutions}
   	\label{fig:image3}
	\end{center}
\end{figure}
\begin{figure}[h!]
	\begin{center}    
    \includegraphics[width=0.7\textwidth]{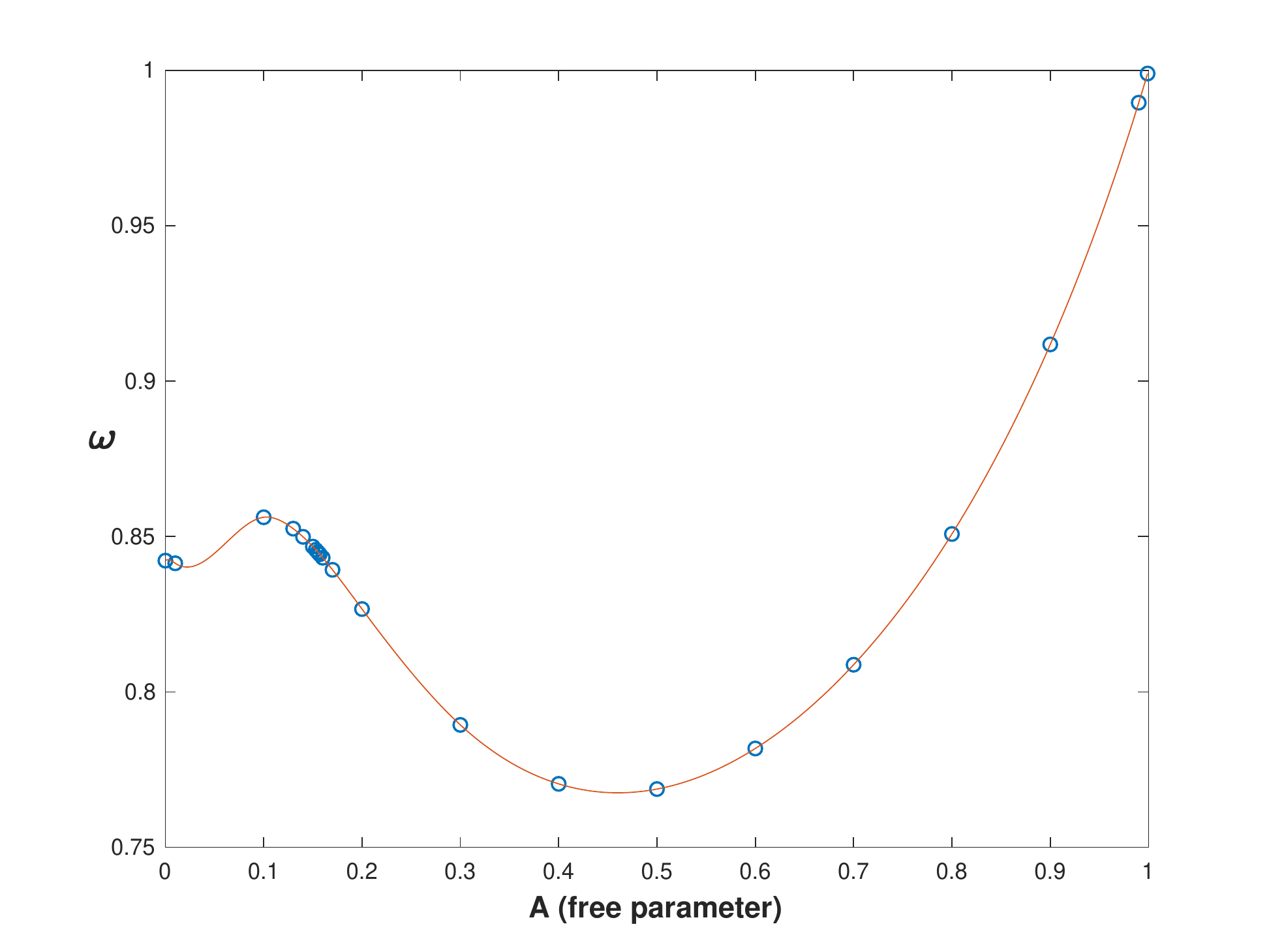}
    \caption{Frequency $\omega$ against the free parameter $A=f(0)$ for the neutral solutions}
   	\label{fig:image3}
	\end{center}
\end{figure}


\clearpage
\section{Conclusion} 

\hspace{1em} We considered a model of 3 minimally coupled canonical fields: electromagnetic, gravitational and massive scalar fields. In the stationary spherically symmetric case we looked for regular, ``soliton-like'' solutions to the system of field equations. All solutions are electrically charged, free of horizon and possess finite positive proper energy associated with inertial (and equal gravitational) mass.

\hspace{1em} Our main goal was to find the solutions whose characteristics reproduce those of a typical elementary particle. Such solutions should correspond to the value $\gamma_0 \sim10^{-40}$ of the sole parameter $\gamma$ which enters the dimensionless Lagrangian, and relates the gravitational/electromagnetic interaction ratio of two charged massive particles.

\hspace{1em} We expected to obtain a proper discrete set of (ground and excited) states making use of the requirement on the electric charge and the mass to achieve minimal values; in this case one could naturally explain the ``magic'' number $\gamma_0$ within the framework of a rather simple and natural 3-field model free of any additionally inserted nonlinearity.

\hspace{1em} Unfortunately, the procedure of numerical integration breaks in the range of parameters $\gamma \le 0.9$, and we were unable to find some analytical explanation of this fact. Thus, it remains an open question whether true {\it particlelike} solutions could exist in the model.

\hspace{1em} Meanwhile, for the range $\gamma \sim 1$ we have followed the dependence of the solutions' characteristics  (charge, mass and frequency) on the other free parameter of the model $B$ and found that, energy indeed has a local minimum at some values of $B$ (both for the ground state and first two excited states, related to the scalar field distributions with nodes). Thus, our model describe a ``semi-quantum'' charged object with a discrete energy/mass spectrum whose dimensions and mass lie in the Planckian range, $10^{-33} cm$ and $10^{-5} gr$, respectively. Appropriate interpretation of such objects is obscure; in literature these are called (regular) ``maximons'' , ``planckeons'',etc.

\hspace{1em} Finally, we considered the limit  $\gamma \to \infty$ in which the electric field is eliminated and one effectively deals with the system of minimally coupled gravitational and massive scalar field. Regular solutions in this model (with and without nonlinear terms in the Klein-Gordon equation)  had been repeatedly examined, starting perhaps from  ~\cite{Kaup}. It is well-known that the metric in such models does not contain horizon, while corresponding mass lie in the solar range. To our knowledge, we are the first to obtain the discrete energy spectrum of such macroscopic objects, from the condition of minimal proper energy (mass). Nonetheless, identification of such objects with real astrophysical entities looks problematic. At the end, ``(mini-) boson stars'' might just be a mathematical construction.

\hspace{1em} We put off for future work further elucidation of the -physically promising but mathematically very complicated- problem of existence of true```particlelike'' solutions (in the range $\gamma <<1$) in the above studied 3-field model. It would be also intriguing to consider the ``$(10^{-40})$'' problem in the framework of the more realistic coupled {\it Maxwell-Einstein-Dirac system of equations}.

\section*{Funding}
The publication has been prepared with the support of the ``RUDN University Program 5-100''.
\section*{Acknowledgments}

The authors are grateful to K. A. Bronnikov for a valuable consultation, to I. Sh. Khasanov for technical support.  
\selectlanguage{English}

\end{document}